\documentclass[authoryear]{elsarticle}
\usepackage{epsfig,graphicx}
\usepackage{graphics}
\usepackage{amsmath}
\usepackage[font=small,labelfont=bf]{caption}

\begin{document}

\begin{frontmatter}

\title{Effects of variability of X-ray binaries on the X-ray luminosity functions of Milky Way}
\author[mymainaddress,mysecondaryaddress]{Nazma Islam \corref{mycorrespondingauthor}}
\cortext[mycorrespondingauthor]{Corresponding author}
\ead{nazma@rri.res.in}
\author[mymainaddress]{Biswajit Paul}
\address[mymainaddress]{Raman Research Institute, C. V. Raman Avenue, Sadashivanagar, Bangalore-560012, India }
\address[mysecondaryaddress]{Joint Astronomy Programme, Indian Institute of Science,
Bangalore-560012, India}

\begin{abstract}
 The X-ray luminosity functions of galaxies have become a useful tool for population studies of X-ray binaries in them. The availability of long term light-curves of X-ray 
binaries with the All Sky X-ray Monitors opens up the possibility of constructing X-ray luminosity functions, by also including the intensity variation effects of the 
galactic X-ray binaries. We have constructed multiple realizations of the X-ray luminosity functions (XLFs) of Milky Way, using the long term light-curves of sources 
obtained in the 2-10 keV energy band with the {\it RXTE}--ASM. The observed spread seen in the value of slope of both HMXB and LMXB XLFs are due to inclusion of 
variable luminosities of X-ray binaries in construction of these XLFs as well as finite sample effects. XLFs constructed for galactic HMXBs in the luminosity 
range $10^{36} - 10^{39}$ erg/sec is described by a power-law model with a mean power-law index of -0.48 and a spread due to variability of HMXBs as 0.19. XLFs constructed 
for galactic LMXBs in the luminosity range $10^{36} - 10^{39}$ erg/sec has a shape of cut-off power-law with mean power-law index of -0.31 and a spread due to variability 
of LMXBs as 0.07.

\end{abstract}


\end{frontmatter}

X-ray emission from a normal galaxy i.e in absence of an AGN or X-ray emitting hot gas, is dominated by collective emission from its X-ray emitting point sources 
\citep{fabbiano2006}. With the advent of {\it Chandra} and {\it XMM}--Newton X-ray telescopes era, detailed study of X-ray emission from nearby galaxies is now 
possible, and has further led to the identification of X-ray binaries in them. This in turn, has helped in construction of X-ray luminosity functions 
(XLFs) of such galaxies. These XLFs can also act as the indicators of star formation rate, stellar mass and evolution of these galaxies 
\citep{grimm2003,ranalli,gilfanov2004,mineo2012,kim2010}. In spite of the high angular resolution and sensitivity of these X-ray telescopes, the large distances to these 
galaxies limits us to probe higher luminosity end of their X-ray binary population. On the contrary, X-ray luminosity functions over a wide luminosities range can be 
constructed with the galactic X-ray binaries, the main hurdle of such an attempt being the distance uncertainity to these galactic sources. 
Initial attempts in constructing Log N-Log S relation for galactic X-ray binaries were made using X-ray sources from {\it Uhuru} catalog \citep{matilsky1973} 
and {\it ASCA} survey \citep{ogasaka1998, sugizaki2001}. \cite{grimm2002} had constructed XLFs for X-ray binaries in Milky Way by averaging over their count-rates 
with the first five years of data of the {\it RXTE} All Sky Monitor. Galactic X-ray luminosity functions are also constructed using 15-50 keV {\it Swift}--BAT 
\citep{voss2010a} and 17-60 keV {\it INTEGRAL} surveys \citep{revnivtsev2008, lutovinov2013}. 
\par
However, the X-ray binaries are highly variable and are often unpredictable. The X-ray binaries show intensity variations by a large factor of a few to 
several orders of magnitude at all timescales above milliseconds. The galactic HMXB population is dominated by Be--HMXBs, in which the accretion onto the compact 
object occurs via formation of equatorial disc by stellar wind \citep{reig2011}. These Be--HMXBs are usually quiescent and occasionally undergo outburst, during which their 
X-ray luminosity increases by several orders of magnitude. Supergiant HMXBs are another class of HMXB population, where the accretion onto the compact object occurs via a 
stellar wind or Roche-lobe overflow of the companion star. The galactic low mass X-ray binaries population are dominated by neutron star LMXBs, where the accretion onto the 
neutron star occurs via Roche Lobe overflow. These systems undergo thermonuclear X-ray bursts lasting few tens to hundreds of seconds \citep{galloway2008}, super-bursts 
lasting up to few hours and outbursts lasting few weeks to few months \citep{campana1998}. The galactic low mass X-ray binaries population also consists of an appreciable 
number of Black hole LMXBs, most of which are transient in nature. During quiescence, they occupy very low luminosity states and occasionally they go into outbursts 
(novae), during which their luminosities increase by several orders of magnitude and they occupy very high luminosity states for a considerable period of 
time \citep{remillard2006}.
\par
Multiple {\it Chandra} and {\it XMM}--Newton observations show that a large fraction of sources in external galaxies are also variable and 
includes many transient sources \citep{voss2007, williams2006}. The X-ray luminosity functions for external galaxies are constructed out of tens of kilosec exposure 
of {\it Chandra} and {\it XMM}--Newton X-ray telescopes, which are essentially snapshot observations of extragalactic X-ray binaries. However, the galactic XLFs constructed 
by averaging luminosities of galactic X-ray binaries over 5 years of {\it RXTE}--ASM observations \citep{grimm2002}, do not represent the true positions of X-ray binaries 
luminosities. It is now possible to construct XLFs of Milky Way, taking into account the variability of X-ray binaries because of the availability of long term light-curves of 
X-ray binaries in our Galaxy with {\it RXTE}--ASM.
\par
The aim of the present work is to construct X-ray luminosity functions of Milky Way taking into account the variable nature of the galactic X-ray binaries. 
We have used 16 years of {\it RXTE}--ASM data to construct differential and integral probability distributions of count-rates for X-ray binaries. 
Using this, we have constructed multiple realizations of X-ray luminosity functions of Milky Way. The X-ray luminosity functions, including completeness corrections for 
flux limited nature of ASM, are constructed separately for HMXBs and LMXBs and are fitted with power-law and power-law with cut-off respectively for each realization. 
These parameter values for multiple realizations are analyzed to estimate the effect of variability of X-ray binaries on the XLFs of Milky Way.

\section{Data and Analysis}
\subsection{RXTE All Sky Monitor light curves}
In order to construct X-ray luminosity functions of Milky Way including the effects of variability of X-ray binaries, long term light-curves of these sources 
are required. The long term light-curves obtained with the {\it RXTE} All Sky Monitor are useful due to all sky nature of its operation and long operational time of 
16 years. It had three coded mask telescopes, was sensitive in 2-10 keV energy band and had a sky coverage of 80$\%$ in  every 90 minutes \citep{levine1996}. 
The sources in the ASM catalogue satisfy the criterion that they have reached an intensity of at least $\>$5 mCrab in the operational time
of {\it RXTE}--ASM and this catalogue excludes some sources like the highly absorbed, hard X-ray sources found in other catalogues like {\it Swift}--BAT, {\it INTEGRAL}--IBIS 
etc. The light-curves of the sources are obtained from {\it RXTE}--ASM archival data.
\footnote{(ftp://heasarc.gsfc.nasa.gov/xte/data/archive/ASMProducts/definitive$\_$1dwell/lightcurves/).}

\subsection{Luminosity functions and Variability effects}

\subsubsection{Construction of differential and integral probability distributions of count-rates}

The cumulative X-ray luminosity function (XLF) for the X-ray binary population in a galaxy is defined as the number distribution N($>L$) of X-ray binaries in a galaxy with 
luminosity greater than L, which is used throughout this paper. The X-ray luminosity functions for the Milky Way are constructed separately for 
84 High Mass X ray Binaries and 116 Low Mass X ray Binaries whose distances are found in literature with reasonable accuracy 
(given in \textbf{Appendix} along with references). 
The light-curves of these sources are extracted from the dwell light-curves and are binned with 1 day bin time.
\par
We quantify the variability in luminosities of each sources in the following way. We first determine the frequency of occurrence of each count rate,
taking into account the errors in the count-rate given in the one day binned light-curve of each source. The errors on the count-rate in {\it RXTE}--ASM 
data consists of counting statistics along with systematic errors. 
For the same value of count-rate of a source, we see different values of errors in the ASM light-curves, which makes it difficult to implement Maximum Likelihood 
method for determination of distribution of sources in presence of systematic errors, in the form suggested by \cite{murdoch1973}.
Though it is not clear if the errors in the ASM light curve are gaussian, in the 
absence of any obvious alternative and for the sake of simplicity, the errors are assumed to be normally distributed for each count-rate. 
The true count-rate of the source corresponding to each data point in the light-curve, is then assumed 
to be normally distributed with the observed count rate as the mean of the distribution and the error associated with the 
count-rate as $\sigma$ of the gaussian distribution. 
\par
To find the probability distribution corresponding to true count-rate, the gaussian distribution of each data 
point are integrated and summed up for all events as given by Equation.\ref{eq:gauss}.\\
\begin{center}
\begin{equation} \label{eq:gauss}
y(a)=\frac{1}{N}\sum_{i=1}^{N}\frac{1}{\sqrt{2\pi}\sigma_{i}}\int^{a+0.05}_{a-0.05} exp[\frac{-(x-c_i)^2}{2\sigma_i^2}] dx
\end{equation}  
\end{center}
where N is the total number of data points in one day binned light-curve, $c_{i}$ is the $i^{th}$ count rate, $\sigma_{i}$ is the error 
associated with each $c_{i}$ and a is the true count-rate bin along x-axis. The bin step for integration is taken to be 0.1 ASM count-rate and 
the integration for every true count-rate is carried out from half of the bin step preceding the count-rate (a-0.05) to half of the bin step after 
the count-rate (a+0.05).
\par
\begin{figure}
\centering
\includegraphics[angle=-90,scale=0.15]{fig1.ps}
\includegraphics[angle=-90,scale=0.15]{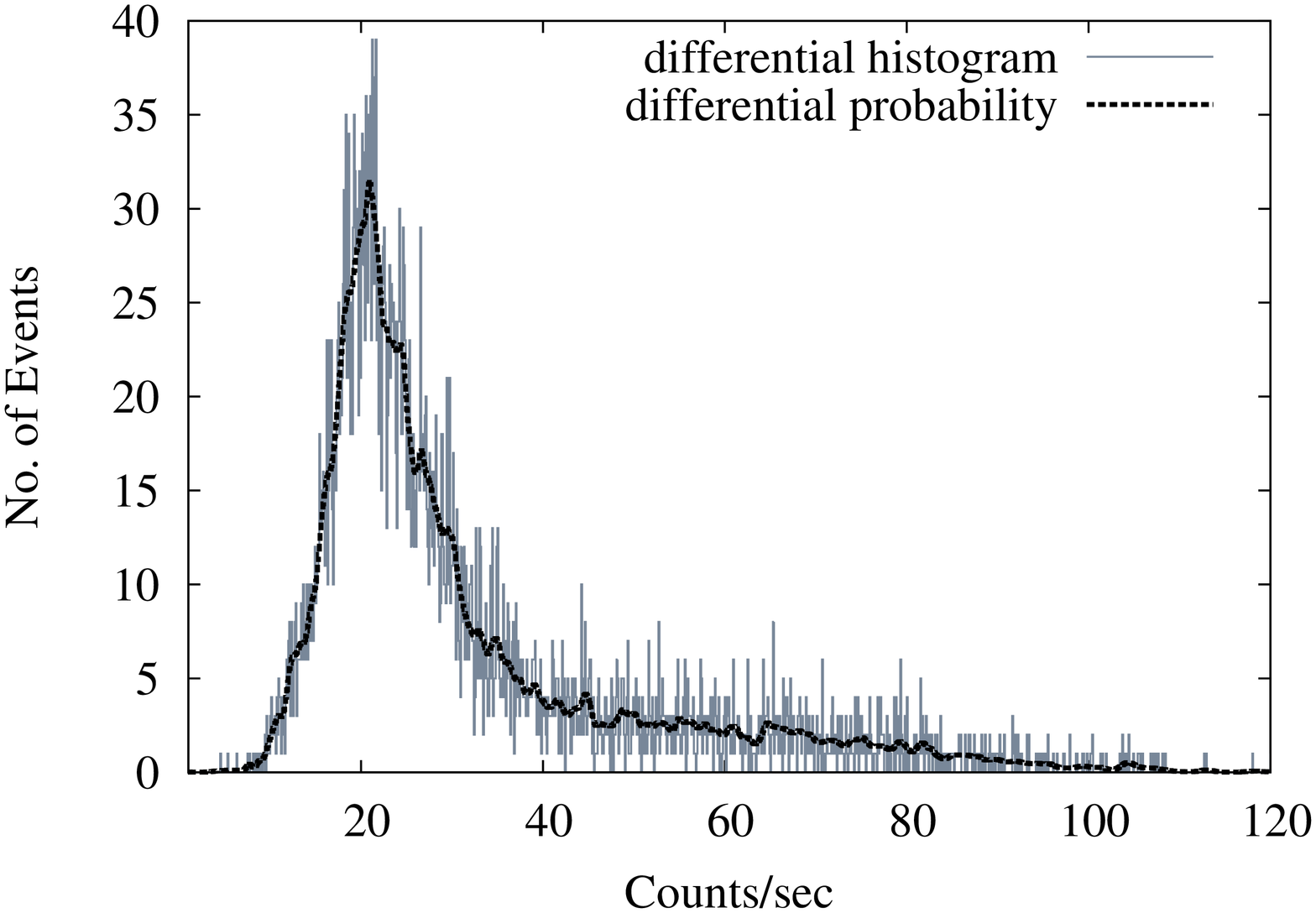}
\includegraphics[angle=-90,scale=0.15]{fig3.ps}
\includegraphics[angle=-90,scale=0.15]{fig4.ps}
\includegraphics[angle=-90,scale=0.15]{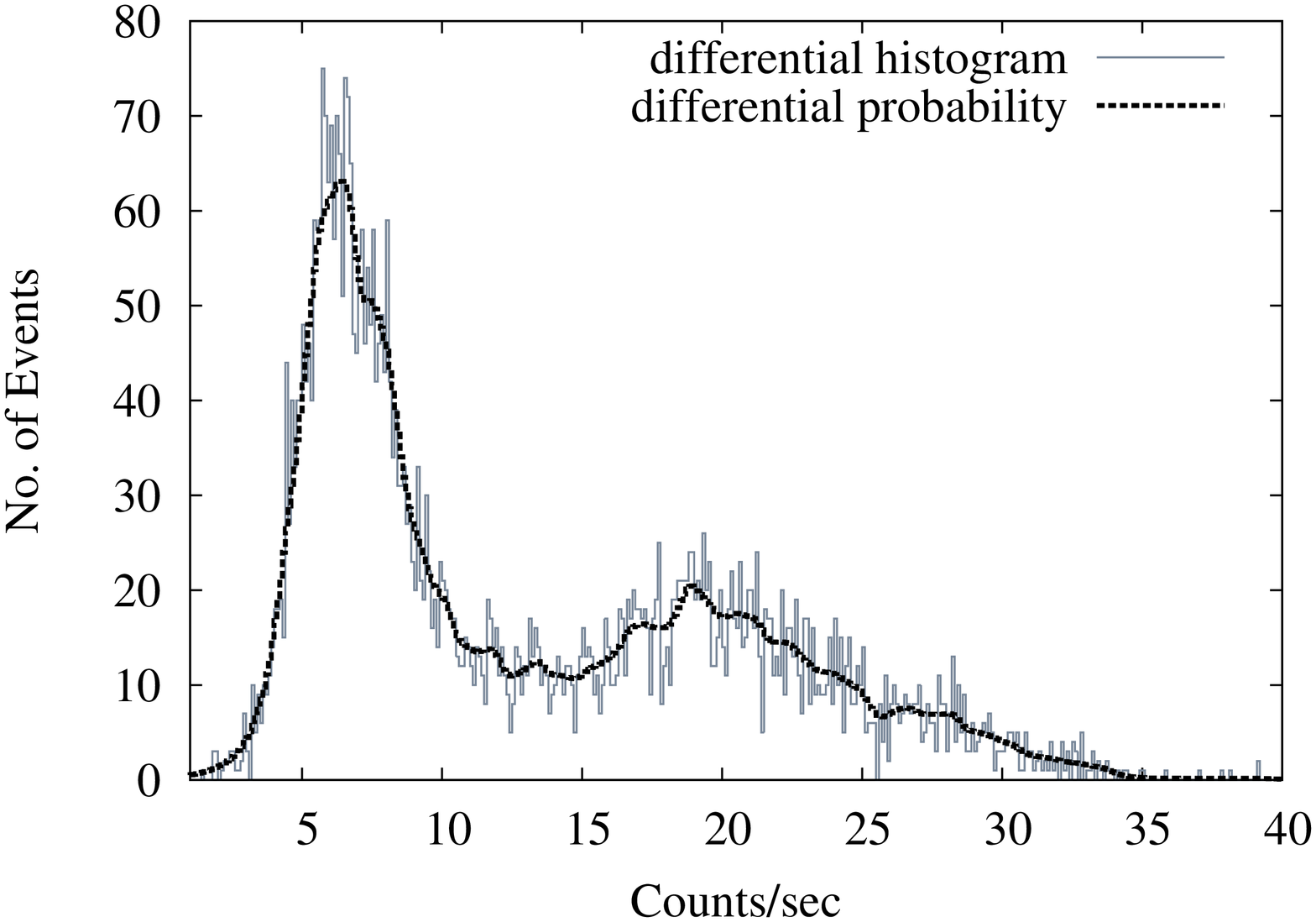}
\includegraphics[angle=-90,scale=0.15]{fig6.ps}
\caption{Light curve of Cyg X--1 (upper left panel) and Cyg X--3 (lower left panel), comparison between differential probability distribution and differential 
histogram (upper middle panel for Cyg X--1 and lower middle panel for Cyg X--3) and integral probability distribution (upper right panel for Cyg X--1 and 
lower right panel for Cyg X--3) are shown here.}
\label{cygx1}
\end{figure}

The probability distribution of count-rates extending below zero obtained by this method are summed up for all data-points and is taken to be zero. 
The events which are registered as NULL in the ASM count rate, are neglected in the analysis. These events indicate the absence of any 
 measurement for example, if the source is close to the Sun.\\
The probability distribution of count-rates constructed by this procedure is called differential probability distribution. 
From differential probability distribution, integral probability distribution is constructed which denotes the probability 
distribution of a source having count-rate greater than the $c_{x}$ (count-rate bin along X axis).\\
The differential and integral probability distributions are constructed for 84 High Mass X ray Binaries and 116 Low Mass X ray Binaries (including field LMXBs 
as well as Globular clusters LMXBs). 
The distributions along with their light curves are shown for Cyg X--1 and Cyg X--3 in Figure.~\ref{cygx1}. 
In Figure.~\ref{cygx1}, the middle panel shows the comparison between differential probability distribution for the Cyg X--1 and Cyg X--3 and 
binned differential histogram of count-rates without accounting for errors on the count-rates.\\

\par
A uniformly distributed random number between 0 and 1 is then compared with the integral probability distribution of each source and the corresponding 
count-rate is selected. The one day binned ASM count rate is converted to flux by assuming a Crab-like spectrum for the sources and using the observed Crab count rate. 
The Crab flux of $2.4\times10^{-8}$ erg $cm^{-2}$ $s^{-1}$ gives a count-rate of 75 counts/sec in 2-10 keV band of {\it RXTE}--ASM and is used for count-rate to flux 
conversion as given in Equation.~\ref{flux} \citep{grimm2002}.\\
\begin{equation}
\label{flux}
F[erg s^{-1} cm^{-2}] = 3.2\times 10^{-10}. R[\mathrm{counts\quad s^{-1}}]
\end{equation}
The flux of the sources determined by Equation.~\ref{flux} are then converted to luminosities with their distances (given in \textbf{Appendix}),
assuming an isotropic emission.\\ 
This process is repeated for each HMXB and LMXB sources and the corresponding selected luminosities are then used in construction of X ray luminosity
distributions for one iteration. The XLFs constructed in this procedure are equivalent to XLFs constructed for snapshot observations of Milky Way 
from the viewpoint of an outside observer.

\subsubsection{Completeness Correction}
Due to the flux limited nature of the ASM sample and the incompleteness in the distance measurements to the X-ray binaries, 
the X-ray luminosity distribution derived in the previous section needs to be corrected for the X-ray binaries not visible to ASM. To account for the 
completeness of the sample, we have used the same model of spatial distribution of X-ray binaries as mentioned in \cite{grimm2002}. 
The HMXB spatial distribution is modeled by disk density distribution parameterized in \cite{dehnen}, with 100\% modulation by galactic spiral arms densities 
(spiral arms computed from \citealt{taylor}). The LMXB spatial distribution is modeled by three component model of \cite{bahcall}, where the parameters of 
bulge, disk and spheroid were appropriately chosen to fit the observed LMXB distribution. The LMXB disk density distribution was 20\% modulated by the 
galactic spiral arm densities. The various parameters of the Galaxy model are tabulated in Table.(4) of \cite{grimm2002}. \\
From Equation.~\ref{completeness} taken from \cite{grimm2002} \\
\begin{equation}
\label{completeness}
(\frac{dN}{dL})_{obs}=(\frac{dN}{dL}) \times (\frac{M( < D(L))}{M_{tot}})
\end{equation}
where $\frac{dN}{dL}$ is the true luminosity function, $(\frac{dN}{dL})_{obs}$ is the flux luminosity distribution constructed from snapshot {\it RXTE}--ASM measurements 
and $(\frac{M( < D(L))}{M_{tot}})$ is the fraction of mass visible to ASM on account of flux limitation and distance incompleteness beyond 10 kpc. 
From Figure.11 of \cite{grimm2002}, which plots the fraction of mass of the galaxy visible to ASM as a function of luminosities, both the LMXB and HMXB mass fraction 
has a flat part of the curve above $10^{36}$ ergs/s. Below $10^{36}$ ergs/s, the mass fraction of both LMXBs and HMXBs visible to ASM rapidly decreases. 
To avoid any artifacts in the snapshot luminosity functions and to ensure that the change in the parameters of the luminosity functions 
is introduced by the luminosity variations in the sources alone, we have constructed the galactic X-ray luminosity functions from $10^{36}$ to $10^{39}$ ergs/sec 
luminosity regime (flat part of the mass fraction of galaxy visible to ASM). 

\subsubsection{Construction of Snapshot Observations}
The randomly selected luminosities of all HMXBs and LMXBs sources in one iteration are used to determine the parameters of X-ray luminosity functions for one 
snapshot observation. Using the same model functions as used in \cite{grimm2002}, the HMXB distribution is then fitted with a power-law in luminosity 
range $10^{36}-10^{39}$ erg/s. \\
\begin{equation}
\label{powerlaw}
N(>L) = K.(\frac{L}{10^{36}})^{-a}
\end{equation}
The LMXB distribution is fitted with a power-law with a cut-off in the luminosity range $10^{36}-10^{39}$ erg/s.\\
\begin{equation}
\label{powerlawc}
N(>L) = K.[(\frac{L}{10^{36}})^{-a}-(\frac{L_{max}}{10^{36}})^{-a}]
\end{equation} 

We calculate the best fit values of the parameters by using the Maximum-Likelihood (ML) method in the form suggested by \cite{crawford}. 
The main advantage of using this implementation of ML is that we use ungrouped data (luminosities) in parameter estimation. 
The value of slope of the power-law (power-law with a cut-off in case of LMXB) is calculated by solving the following equation \citep{crawford}
\begin{equation}
 \label{mle}
 \frac{M}{a} - \sum_{i} \mathrm{ln} s_{i} - \frac{M \mathrm{ln} b}{b^{a}-1} = 0
\end{equation}
where M is the total number of HMXB or LMXB sources in a snapshot observations having luminosity greater than $10^{36}$ ergs/sec, {\it s}$_{i}$ is the luminosities of 
each source in units of $10^{36}$ and b (in units of $10^{36}$) is the maximum value of luminosity present in each iteration. The statistical error on slope {\it a} is 
\citep{crawford}
\begin{equation}
 \label{mle_sigma}
\Sigma_{a} = \frac{a}{\sqrt M}(1-\frac{a^{2}(\mathrm{ln}b)^{2}}{b^{a}})^{-\frac{1}{2}}
\end{equation}

\begin{figure}
\centering
\includegraphics[angle=-90,scale=0.2]{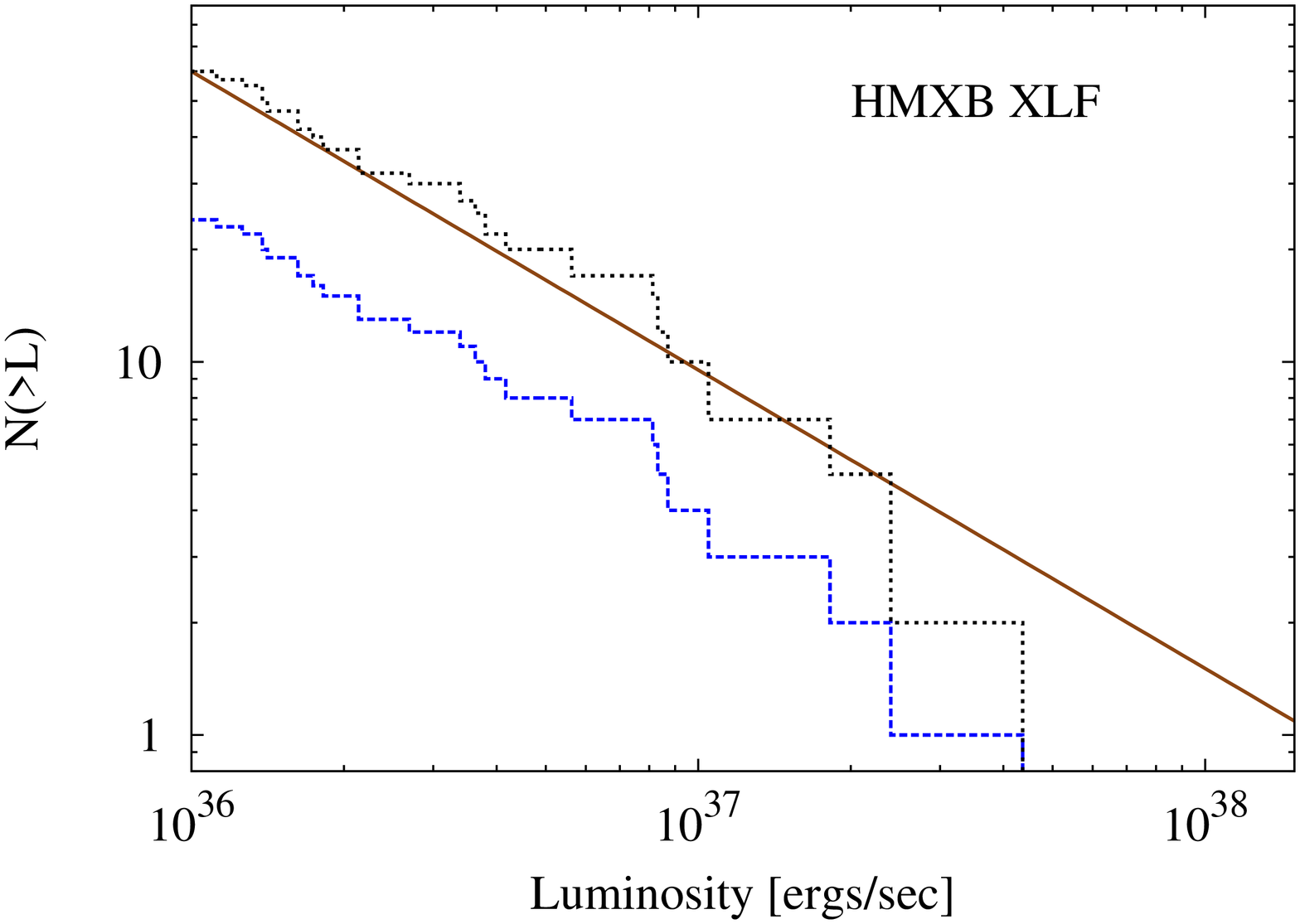}
\includegraphics[angle=-90,scale=0.2]{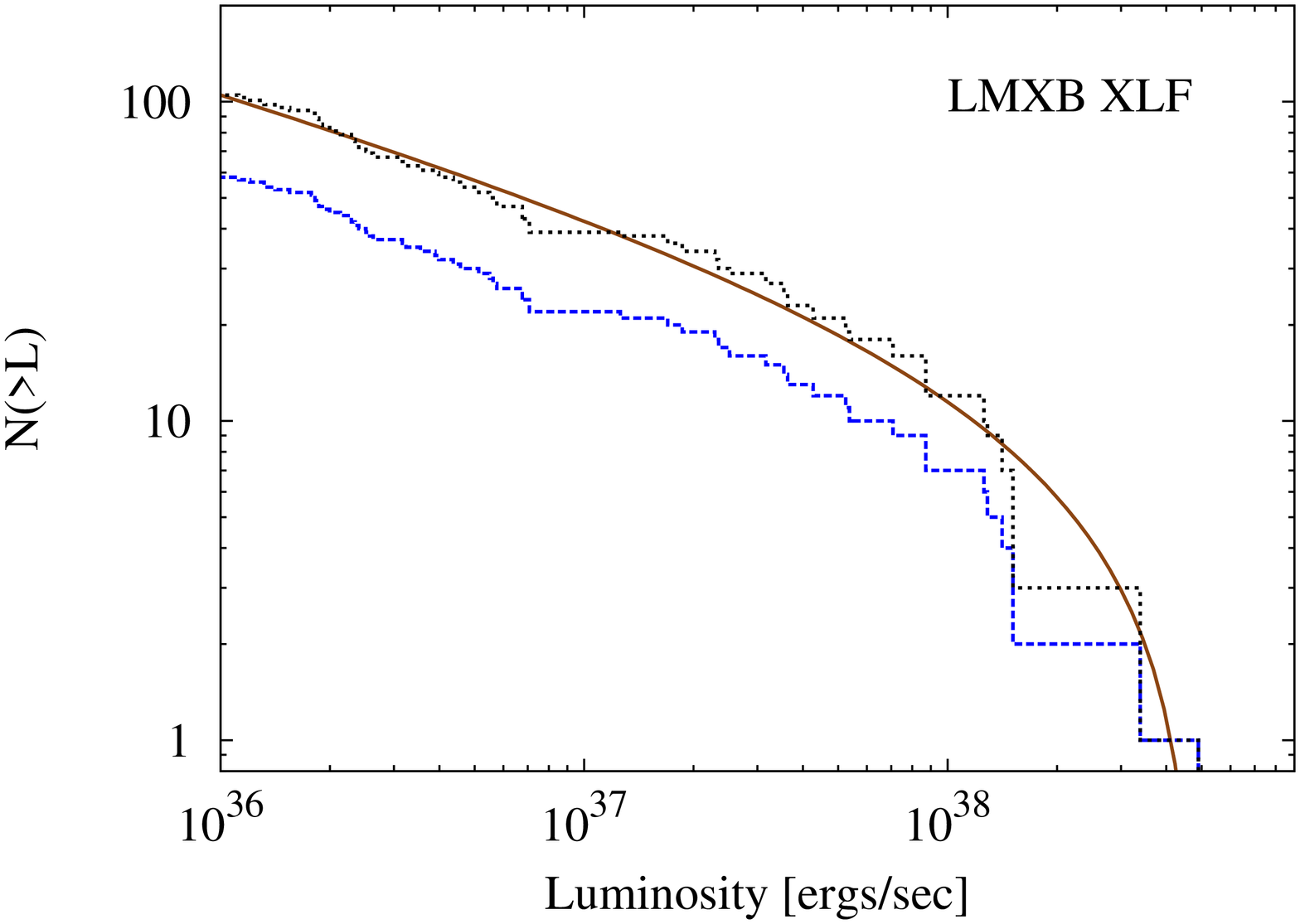}
\caption{The observed (dashed line) and volume corrected (dotted line) Log N-Log L distribution of HMXB (left panel) and LMXB (right panel) for 
one arbitrarily chosen iteration. Solid line represents the best fit to the distribution with the parameter values obtained from M-L method, 
power-law model given by Equation.~\ref{powerlaw} for HMXBs and power-law with a cut-off model given by Equation.~\ref{powerlawc} for LMXBs.}
\label{hmxbfit}
\end{figure}

As mentioned in \cite{grimm2002}, the cut-off of the LMXB distribution is taken as the maximum luminosity of the sample 
in a given iteration. The normalisation K of the HMXB XLFs for a given iteration is taken to be the number of sources (including completeness corrections 
obtained from Figure.11 of \cite{grimm2002}) having luminosities greater than $10^{36}$ ergs/sec. For the LMXB XLFs, it is the number of sources (including completeness 
corrections obtained from Figure.11 of \cite{grimm2002}) greater than $10^{36}$ ergs/sec divided by $\mathrm{(1-L_{max}^{-a})}$. 
A plot of the XLF and the best fit model for one arbitrarily chosen iteration, without and with completeness corrections, 
is shown for HMXBs and LMXBs in Figure.~\ref{hmxbfit}.

\par
Using different seeds in the random number generator, the above process is repeated 10,000 times and for each iteration,
 different values of slopes are obtained from Equation.~\ref{mle} along with its statistical error. The parameter values obtained for different iterations are plotted as 
 histograms. The mean value and standard deviation ($\sigma$) of parameters are calculated from their respective histograms. 
The mean values of parameters along with their $\sigma$ are tabulated in Table.~\ref{table1} and the histogram of parameters are shown in 
Figure.~\ref{hmxbparameter} for LMXB XLFs and HMXB XLFs respectively. We have also calculated the mean of statistical error $\Sigma_{a}$ on the values of slopes for each 
iteration.

\par
To determine the goodness of fit of models given by Equation.~\ref{powerlaw} to the HMXB XLFs and Equation.~\ref{powerlawc} to the LMXB XLFs, 
we have performed Kolmogorov-Smirnoff test (KS test) on the data and compared it with its respective model. 
The average KS probability ({\it p}) value for 10000 iterations of HMXB XLFs is $\sim$ 0.44 and the average {\it p} value for 
10000 iterations of LMXB XLFs is $\sim$ 0.41. Out of the total 10000 iterations for each type, 9359 of the HMXB XLFs and
9548 of the LMXB XLFs have {\it p} $>$ 0.1.

\section{Discussions and Conclusions}

Previous work by \cite{grimm2002} had utilized the first 5 years of {\it RXTE}--ASM data for constructing the averaged XLFs separately using 25 HMXBs and 84 LMXBs. 
However, these galactic XLFs do not include the variability effects of X-ray binaries and for a transient X-ray binary, the averaged count-rates do not represent the true 
positions of such systems in the XLF. Therefore, we have used 16 years light-curves of 2-10 keV energy band of {\it RXTE}--ASM of 84 HMXBs and 116 LMXBs to construct 
multiple realizations of the XLFs for the Milky Way by incorporating the variable nature of X-ray binaries. The differential and integral probability distributions of 
count-rates in 2-10 keV {\it RXTE}--ASM light-curves are constructed for each source. The snapshot luminosity distribution are constructed separately for galactic HMXBs and 
LMXBs and includes completeness corrections for flux limited nature of ASM sample. For each iteration of the HMXBs, the XLF is fitted with a power-law, whereas for the LMXBs, 
the XLF is fitted with a power-law with cut-off. The value of slope of the power-law and power-law with cut-off is estimated from Maximum Likelihood analysis, along with the 
statistical error on the value of slope. This process is repeated for 10,000 realizations of XLFs of Milky Way and obtain values of XLFs parameters for all these realizations. 
The mean and $\sigma$ of each parameter is estimated from these values of 10,000 realizations. For each realization, we also obtain the statistical error on the value of 
slope $\Sigma_{a}$, which is then averaged.
\par
The HMXB luminosity function taking into account variability of High Mass X-ray binaries, in luminosity range $10^{36}-10^{39}$ erg/s is given by:\\
\begin{equation}
\label{hmxblf}
N(>L) = (54\pm8)(\frac{L}{10^{36}erg/s})^{-(0.48\pm0.19)}
\end{equation}
The LMXB luminosity function taking into account the variability of Low Mass X-ray binaries, in luminosity range $10^{36}-10^{39}$ erg/s is given by:\\
\begin{equation}
\label{lmxblf}
N(>L) = (127\pm8)((\frac{L}{10^{36}erg/s})^{-(0.31\pm0.07)}-((4.28\pm1.6)\times10^{3})^{-(0.31\pm0.07)})
\end{equation}

\par
The value of $\sigma$ quoted here as the spread in the value of slope for the HMXB and LMXB luminosity functions is mainly due to the variations in the XLFs arising 
from the variability effects of X-ray binaries. This is different from the statistical error on the value of slope $\Sigma_{a}$, given in Equation.~\ref{mle_sigma}, 
arising due to finite N sample effects. As seen in Table.1, the spread in the slope seen due to variability of X-ray binaries is marginally larger than the averaged 
$\Sigma_{a}$ for both HMXB and LMXB XLFs. 

\par
For LMXB distribution, the most luminous source in X-rays for a given iteration determines the cut-off of the distribution. 
The most luminous and persistent LMXB X-ray source in 2-10 keV energy band of {\it RXTE}--ASM in our Galaxy is Sco X--1, which determines 
the cut-off of LMXB distribution for a majority of the snapshots.
However, some of the galactic LMXB transients (like Aql X--1 and other Black hole binaries like GRS 1915+105) during outbursts occasionally outshines 
Sco X--1, which in turn influences the cut-off of the LMXB distribution for the snapshots in which such luminosities are included. 
As shown in Table.~\ref{table1}, the cut-off of the LMXB distribution is of the order of Eddington luminosity for 1.4 $M_{\odot}$ He accreting neutron star LMXB.\\ 
The slope of the LMXB XLF is similar to that found in previous study by \cite{grimm2002}. However, the slope of the HMXB XLF is smaller than that of HMXB XLF 
given in \cite{grimm2002}. This discrepancy could be due to inclusion of more Be--HMXBs in the sample, which have high luminosity during the outbursts 
and are non-detectable with ASM during their quiescence. Also HMXBs like LSI +61303, X Per and Supergiant Fast X-ray Transients like IGR J18483--0311, 
are included in some snapshot observations when they are in high luminosity states. 

\par
\begin{figure}
\includegraphics[angle=-90,scale=0.15]{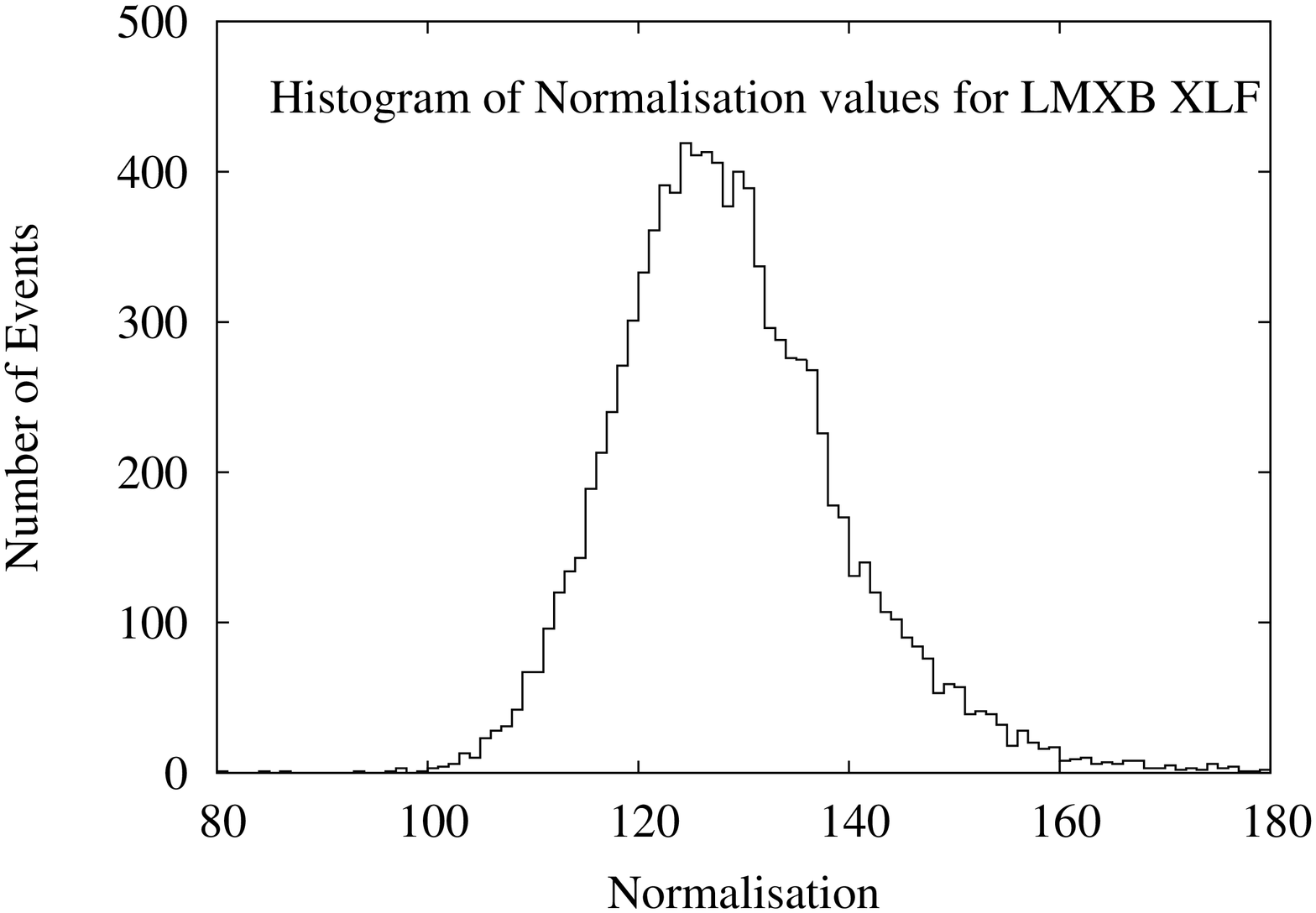}
\includegraphics[angle=-90,scale=0.15]{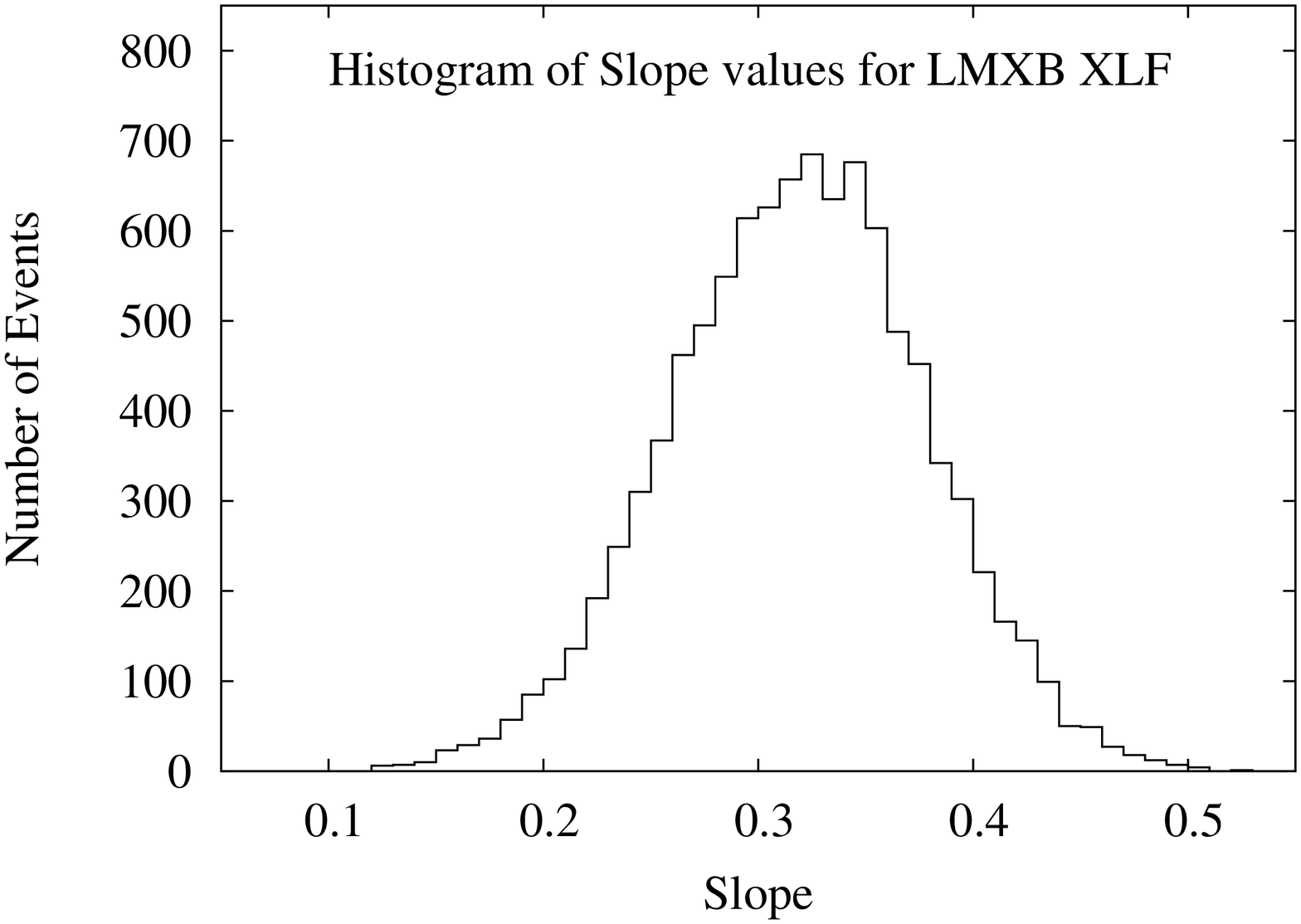}
\includegraphics[angle=-90,scale=0.15]{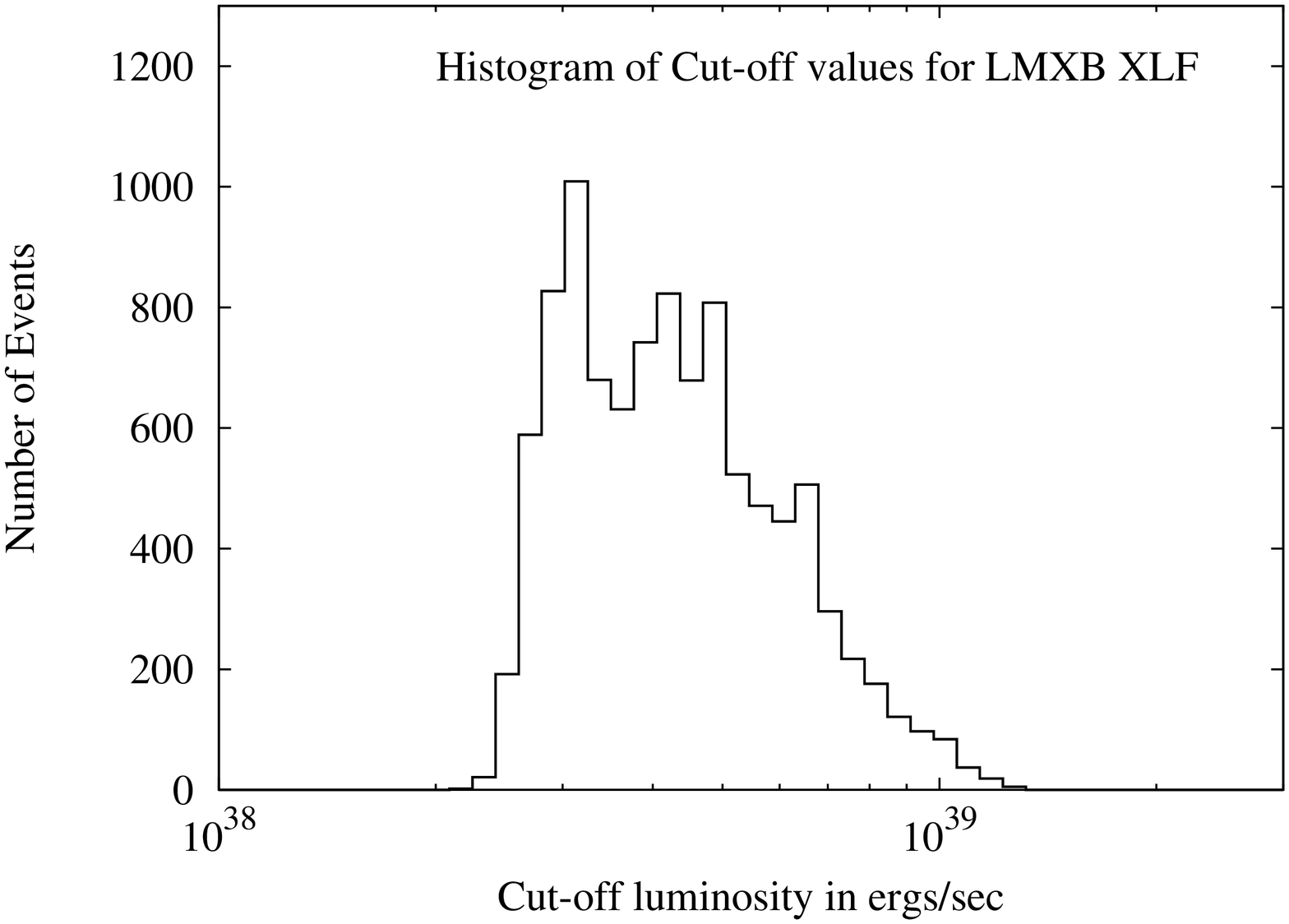}

\includegraphics[angle=-90,scale=0.15]{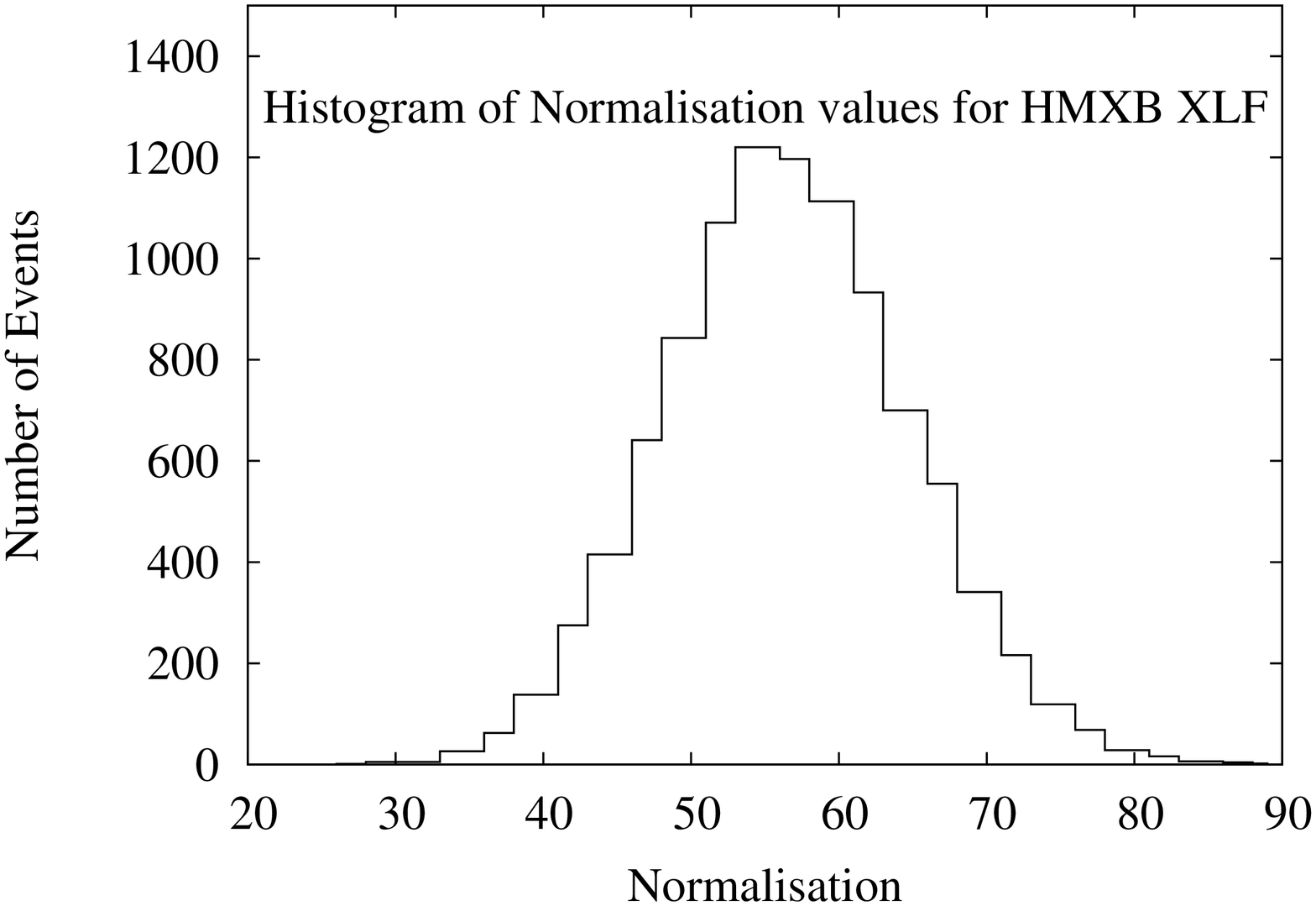}
\includegraphics[angle=-90,scale=0.15]{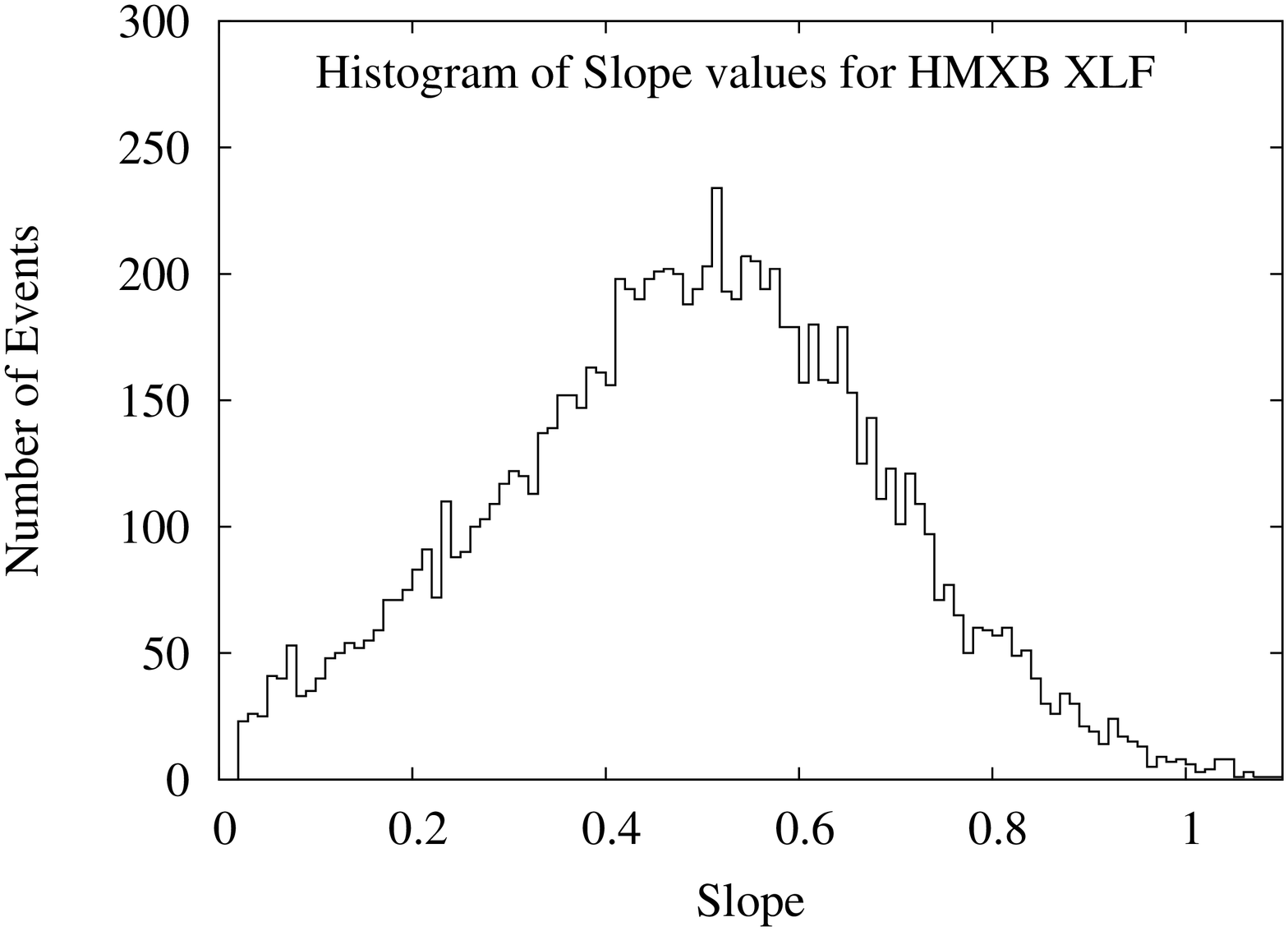}

\caption{Histograms of LMXB XLFs and HMXB XLFs parameter values for 10,000 iterations. 
Top panel shows the distribution of Normalisation (left panel),
Slope (middle panel) and Cut-off (right panel) of LMXB XLFs. 
Lower panel shows the distribution of Normalisation (left panel) and 
Slope (right panel) of HMXB XLFs.}
\label{hmxbparameter}
\end{figure}

From Figure.~\ref{hmxbparameter}, we see that the slope histogram of HMXB XLFs have a larger spread than the slope histogram of LMXB XLFs. 
Since two-thirds of the galactic HMXB population consists of Be-HMXBs and HMXB XLFs are mostly influenced by the collective emission properties of these 
stellar wind driven Be-HMXB systems, we can infer that a large fraction of transients in the underlying population of HMXBs leads to larger spread in HMXB XLFs 
than that seen in LMXB XLFs. \cite{postnov2003} and \cite{bhadkamkar2012} have shown that the power law shape of the HMXB XLF can be explained by the stellar wind properties 
of massive stars. Most of the HMXBs systems are NS-HMXBs, with only 6 HMXB having black holes. Cyg X--1 and Cyg X--3 are the most luminous HMXB systems and therefore always 
occupy higher end of HMXB XLFs.

\par
The LMXB population consists of both field LMXBs as well as Globular Clusters (GCs) LMXBs. The LMXBs in GCs have different XLF behaviour compared to 
field LMXBs as seen in the study of LMXBs in the bulge of M 31 \citep{voss2007}. Since there are only 12 LMXBs in the catalogue of galactic LMXBs \citep{liu2007}, 
we have ignored the possible effects of LMXBs in GCs on the LMXB XLFs. The LMXB XLFs are mostly influenced by the collective emission properties of these 
NS-LMXBs with the break in the XLF either due to the change in the mass transfer rate in the binary systems \citep{postnov2005} or due to fraction of 
Giant donors in population of LMXBs \citep{revnivtsev2011}. There are also an appreciable number of transient BH LMXBs, which are either in quiescence and occupy very low 
luminosity states or undergo outbursts for a brief period of time and occupy high luminosity end of XLFs. 

\par
In this paper, we have probed one aspect of probable variation in XLFs due to variable nature of X-ray binaries. In case of star-bursts galaxies, the number of HMXBs present 
in a galaxy is related to its Star Formation Rate (SFR), whereas for elliptical galaxies, the number of LMXBs present in a galaxy is related to its 
stellar mass \citep{mineo2012}. It will be interesting to probe the variation in extra-galactic XLFs due to both variability of X-ray binaries as well as the number of 
X-ray binaries present, which differs from galaxy to galaxy. 

\vspace{10mm}

\textbf{ACKNOWLEDGMENT}\\
This research has made use of quick-look results of {\it RXTE}--ASM obtained through High Energy Astrophysics Science
Archive Research Center Online Service (HEASARC), provided by the NASA/Goddard Space Flight Center. We thank Shiv Sethi and Harshal Bhadkamkar for 
useful discussions. 

\bibliographystyle{elsarticle-harv}

\newpage

\begin{table}
\caption{Mean parameters values with their $\sigma$ for 10,000 iterations. $\Sigma_{a}$ is the averaged statistical error on the value of slope {\it a}}
\label{table1}
\centering
\begin{tabular}{| c | c | c | c | c | c | c | c |}
\hline
Sample & a & $\sigma_{a}$ & $\Sigma_{a}$ & K & $\sigma_{K}$ & L$_{max}$ & $\sigma_{L_{max}}$\\
 & & & & & & ($10^{38}$ ergs/sec) & ($10^{38}$ ergs/sec)\\
 \hline
HMXB & 0.48 & 0.19 & 0.11 & 54 & 8 & - & -\\
\hline
LMXB & 0.31 & 0.07 & 0.05 & 127 & 8 & 4.28 & 1.6\\
\hline
\end{tabular}
\end{table}

\begin{table}
\footnotesize
\caption*{\textbf{Appendix: Distances to X ray binaries}}
\label{appendix}
\centering
\begin{tabular}{|c c c c|}
\hline
X ray sources & Distance(Kpc) & Type & References\\
\hline
AqlX-1  & 3.4     & LMXB	& \cite{paradijs} \\               
CirX-1  & 10.9   & LMXB    & \cite{paradijs} \\             
CygX-1  & 2.1    & HMXB	& \cite{grimm2002}   \\             
CygX-2  & 11.3   & LMXB & \cite{paradijs}   \\             
CygX-3  & 9.0    & HMXB	& \cite{paradijs}   \\             
EXO0748-676     & 7.6    & LMXB	& \cite{galloway2008} \\        
GROJ1744-28     & 8.5    & LMXB & \cite{paradijs}  \\      
GRS1915+105     & 12.5   & LMXB	& \cite{grimm2002}  \\      
GS1124-684      & 5.5    & LMXB	& \cite{paradijs}  \\      
GS1843+009      & 12.5   & HMXB & \cite{grimm2002} \\       
GS2000+250      & 2.7   & LMXB & \cite{liu2007}  \\      
GS2023+338      & 4.3    & LMXB	& \cite{paradijs} \\       
GX5-1           & 7.2    & LMXB & \cite{grimm2002} \\       
GX13+1         & 7.0    & LMXB	& \cite{grimm2002} \\       
GX17+2         & 9.5   & LMXB & \cite{galloway2008} \\       
GX301-2        & 3.1    & HMXB & \cite{coleiro2012} \\       
GX340+0        & 11.0   & LMXB & \cite{paradijs} \\       
GX349+2        & 9.2    & LMXB & \cite{grimm2002} \\       
GX354+0        & 5.7    & HMXB & \cite{grimm2002} \\       
KS1731-260     & 5.6    & LMXB & \cite{galloway2008}\\        
SerX-1         & 8.4    & LMXB & \cite{grimm2002}  \\      
V4641Sgr       & 9.9    & HMXB & \cite{grimm2002} \\       
X1538-522      & 4.5   & HMXB & \cite{bodaghee} \\       
X1608-522      & 4.0   & LMXB & \cite{paradijs}\\        
X1624-490      & 13.5  & LMXB & \cite{grimm2002} \\       
X1636-536      & 5.95   & LMXB & \cite{galloway2008} \\       
X1657-415      & 7.1   & HMXB & \cite{bodaghee} \\       
X1705-440      & 7.4   & LMXB & \cite{grimm2002} \\       
X1715-321      & 6.1    & HMXB & \cite{grimm2002}\\        
X1735-444      & 6.5    & LMXB & \cite{galloway2008}\\ 
X1812-121      & 3.8    & HMXB & \cite{grimm2002}\\        
X1820-303      & 4.94    & LMXB & \cite{galloway2008}\\        
X1905+000      & 7.7    & HMXB & \cite{grimm2002}\\        
X1908+075      & 6.4    & HMXB & \cite{grimm2002}\\        
X1916-053      & 6.8    & LMXB & \cite{galloway2008}\\        
XTEJ1550-564   & 5.3    & LMXB & \cite{grimm2002}\\        
X0115+634      & 5.3    & HMXB & \cite{coleiro2012}\\        
RXJ0146.9+6121 & 2.3    & HMXB & \cite{negueruela}\\        
V0332+53       & 7.0   & HMXB & \cite{negueruela} \\       
X0535+262      & 3.8   & HMXB & \cite{coleiro2012} \\       
X0726-260      & 5.0    & HMXB & \cite{coleiro2012} \\       
GROJ1008-57    & 5.0    & HMXB & \cite{negueruela} \\       
X1118-616      & 3.2    & HMXB & \cite{coleiro2012} \\       
X1145-619      & 4.3    & HMXB & \cite{coleiro2012} \\       
X1417-624      & 7.0    & HMXB & \cite{coleiro2012}  \\
EXO2030+375    & 3.1    & HMXB & \cite{coleiro2012} \\       
CepX-4         & 3.7    & HMXB & \cite{coleiro2012}  \\      
ScoX-1         & 2.8    & LMXB & \cite{grimm2002} \\      
X0836-426      & 8.2    & LMXB & \cite{galloway2008} \\       
X0918-548      & 4.0    & LMXB & \cite{galloway2008} \\      
X1254-690      & 15.5   & LMXB & \cite{galloway2008} \\       
X1323-619      & 11.0   & LMXB & \cite{galloway2008} \\      
X1702-429      & 4.19   & LMXB & \cite{galloway2008} \\       
XTEJ1723-376   & 10.0   & LMXB & \cite{galloway2008}\\        
X1724-307      & 5.0    & LMXB & \cite{galloway2008} \\      
SL1735-269     & 5.6    & LMXB & \cite{galloway2008}\\

\hline
\end{tabular}
\end{table}

\begin{table}
 \footnotesize
\caption*{\textbf{Appendix: Distances to X ray binaries(continued)}}
\centering
\begin{tabular}{|c c c c|}
\hline
X ray sources & Distance & Type & References\\
\hline 
XTEJ1739-285   & 7.3    & LMXB & \cite{galloway2008} \\       
SAXJ1747-2853   & 5.2    & LMXB & \cite{galloway2008} \\       
IGRJ17473-2721 & 4.9    & LMXB & \cite{galloway2008} \\
SL1744-300     & 8.4   & LMXB & \cite{galloway2008} \\   
GX3+1          & 5.0    & LMXB & \cite{galloway2008} \\ 
X1744-361       & 8.4    & LMXB & \cite{galloway2008} \\       
EXO1745-248    & 4.73   & LMXB & \cite{galloway2008} \\      
X1746-370      & 16.0   & LMXB & \cite{galloway2008}\\        
SAXJ1750.8-2900& 5.21    & LMXB & \cite{galloway2008}\\        
GRS1747-312    & 9.0    & LMXB & \cite{galloway2008} \\       
XTEJ1759-220   & 16.0   & LMXB & \cite{galloway2008}  \\      
SAXJ1808.4-3658& 2.77   & LMXB & \cite{galloway2008} \\       
XTEJ1814-338   & 7.9    & LMXB & \cite{galloway2008}  \\      
GS1826-238     & 6.7    & LMXB & \cite{galloway2008}   \\ 
X1832-330      & 6.7    & LMXB & \cite{galloway2008} \\       
HETEJ1900.1-2455        & 3.6    & LMXB & \cite{galloway2008}\\
XTEJ2123-058            & 14.0   & LMXB & \cite{galloway2008}\\
IGRJ18027-2017         & 12.4   & HMXB & \cite{torrejon}\\
SAXJ1818.6-1703         & 2.7   & HMXB & \cite{coleiro2012}\\
IGRJ18483-0311          & 2.83   & HMXB & \cite{torrejon}\\
IGRJ1914+0951          & 3.6    & HMXB & \cite{torrejon}\\
GROJ0422+32            & 2.5    & LMXB & \cite{jonker}\\
X0620-003               & 1.2    & LMXB & \cite{jonker}\\
GRS1009-45             & 5.7    & LMXB & \cite{jonker}\\
XTEJ1118+480           & 1.8    & LMXB & \cite{jonker}\\
X1543-475              & 7.5    & LMXB & \cite{jonker}\\
GROJ1655-40            & 3.2    & LMXB & \cite{jonker}\\
GX339-4                & 6.0    & LMXB & \cite{jonker}\\
X1705-250              & 8.6    & LMXB & \cite{jonker}\\
XTEJ1859+226           & 6.3    & LMXB & \cite{jonker}\\
X1658-298              & 8.4    & LMXB & \cite{jonker}\\
SAXJ1712.6-3739        & 5.9    & LMXB & \cite{jonker}\\
RXJ1718.4-4029         & 6.4    & LMXB & \cite{jonker}\\
SAXJ1810.8-2609        & 5.1   & LMXB & \cite{jonker}\\
X0656-072              & 3.9    & HMXB & \cite{grimm2002}\\
X1354-644              & 27.0   & LMXB & \cite{wu}\\
XTEJ1650-500           & 2.6    & LMXB & \cite{wu}\\
IGRJ00291+5934         & 2.6    & LMXB & \cite{wu}\\
XTEJ0929-314           & 10.0   & LMXB & \cite{wu}\\
XTEJ1751-305           & 8.5    & LMXB & \cite{wu}\\
X0114+650              & 6.5    & HMXB & \cite{coleiro2012}\\
X1845-024              & 10.0   & HMXB & \cite{bodaghee}\\
X2206+543              & 3.4    & HMXB & \cite{coleiro2012}\\
X1700-377              & 1.8    & HMXB & \cite{coleiro2012}\\
AXJ1820.5-1434         & 8.2    & HMXB & \cite{bodaghee}\\
GammaCas               & 0.17   & HMXB & \cite{coleiro2012}\\
GX304-1                & 1.3    & HMXB & \cite{coleiro2012}\\
X1907+097              & 5.0    & HMXB & \cite{bodaghee}\\
RXJ0037.2+6121         & 3.0    & HMXB & \cite{bodaghee}\\
IGRJ01363+6610         & 2.0    & HMXB & \cite{bodaghee}\\
IGRJ01583+6713         & 4.1    & HMXB & \cite{coleiro2012}\\
IGRJ06074+2205         & 4.5    & HMXB & \cite{coleiro2012}\\
IGRJ08408-4503         & 3.4    & HMXB & \cite{coleiro2012}\\
IGRJ11215-5952         & 7.3    & HMXB & \cite{coleiro2012}\\
IGRJ11305-6252         & 3.6    & HMXB & \cite{coleiro2012}\\
IGRJ11435-6109         & 9.8    & HMXB & \cite{coleiro2012}\\

\hline
\end{tabular}
\end{table}

\begin{table}
 \footnotesize
\caption*{\textbf{Appendix: Distances to X ray binaries(continued)}}
\centering
\begin{tabular}{|c c c c|}
\hline
X ray sources & Distance & Type & References\\
\hline 
IGRJ16195-4945         & 4.5    & HMXB & \cite{bodaghee}\\
IGRJ16318-4848         & 1.6    & HMXB & \cite{bodaghee}\\
IGRJ16320-4751         & 3.5    & HMXB & \cite{bodaghee}\\
IGRJ16393-4643         & 10.6   & HMXB & \cite{bodaghee}\\
IGRJ16418-4532         & 13.0   & HMXB & \cite{bodaghee}\\
IGRJ16479-4514         & 2.8    & HMXB & \cite{bodaghee}\\
IGRJ17544-2619         & 3.2    & HMXB & \cite{bodaghee}\\
IGRJ18450-0435         & 6.4    & HMXB & \cite{coleiro2012}\\
KS1947+300             & 8.5    & HMXB & \cite{coleiro2012}\\
PSR1259-63             & 1.7   & HMXB & \cite{coleiro2012}\\
RXJ1826.2-1450         & 2.5    & HMXB & \cite{bodaghee}\\
SAXJ2103.5+4545        & 8.0    & HMXB & \cite{coleiro2012}\\
SS433                  & 5.5    & HMXB & \cite{bodaghee}\\
SWIFTJ2000.6+3210      & 8.0    & HMXB & \cite{bodaghee}\\
VelaX-1                & 2.2    & HMXB & \cite{coleiro2012}\\
XPer                   & 1.2  & HMXB & \cite{coleiro2012}\\
XTEJ1543-568           & 10.0   & HMXB & \cite{bodaghee}\\
XTEJ1810-189           & 11.5   & HMXB & \cite{bodaghee}\\
XTEJ1829-098           & 10.0   & HMXB & \cite{bodaghee}\\
XTEJ1855-026           & 10.8   & HMXB & \cite{coleiro2012}\\
X1145-616              & 8.5    & HMXB & \cite{bodaghee}\\
CenX-3                 & 9.0    & HMXB & \cite{grimm2002}\\
CenX-4                 & 1.6    & HMXB & \cite{grimm2002}\\
GRS1716-249            & 2.4    & LMXB & \cite{liu2001}\\
HK1732-304             & 5.2    & LMXB & \cite{liu2001}\\
1E1024.1-5733          & 3.0    & HMXB & \cite{grimm2002}\\
EXO1846-031            & 7.0    & LMXB & \cite{grimm2002}\\
GX9+1                  & 7.2    & LMXB & \cite{grimm2002}\\
GX9+9                  & 7.0    & LMXB & \cite{grimm2002}\\
LSI+61303              & 2.3    & HMXB & \cite{grimm2002}\\
RXJ0812.4-3115         & 8.6    & HMXB & \cite{coleiro2012}\\
RXJ1037.5-5648         & 5.0    & HMXB & \cite{grimm2002}\\
SctX-1                 & 10.0   & HMXB & \cite{grimm2002}\\
X1630-472              & 4.0    & LMXB & \cite{grimm2002}\\
X0042+327              & 7.0    & LMXB & \cite{grimm2002}\\
X0142+614              & 1.0    & LMXB & \cite{grimm2002}\\
X0614+091              & 3.0    & LMXB & \cite{grimm2002}\\
X1543-624              & 10.0   & LMXB & \cite{grimm2002}\\
X1553-542              & 10.0   & HMXB & \cite{grimm2002}\\
X1556-605              & 10.0   & LMXB & \cite{grimm2002}\\
X1627-673              & 8.0    & LMXB & \cite{grimm2002}\\
X1730-333              & 8.0    & LMXB & \cite{grimm2002}\\
X1755-338              & 6.0    & LMXB & \cite{grimm2002}\\
X1803-245              & 8.0    & LMXB & \cite{grimm2002}\\
X1822-000              & 4.0    & LMXB & \cite{grimm2002}\\
X1822-371              & 2.5    & LMXB & \cite{grimm2002}\\
X1957+115              & 7.0    & LMXB & \cite{grimm2002}\\
X2129+470              & 2.2    & LMXB & \cite{grimm2002}\\
XTEJ0421+560           & 1.0    & HMXB & \cite{grimm2002}\\
1E1048.1-5937          & 3.0    & HMXB & \cite{grimm2002}\\
1E1740.7-2942          & 8.5    & LMXB & \cite{grimm2002}\\
1E2259.0+5836          & 4.0    & LMXB & \cite{grimm2002}\\
GCX-1                  & 8.5    & LMXB & \cite{grimm2002}\\
GS0834-430             & 7.1    & HMXB & \cite{coleiro2012}\\
GRS1739-278            & 6.0    & LMXB & \cite{grimm2002}\\
GRS1758-258            & 8.5    & LMXB & \cite{grimm2002}\\
\hline
\end{tabular}
\end{table}

\begin{table}
 \footnotesize
\caption*{\textbf{Appendix: Distances to X ray binaries(continued)}}
\centering
\begin{tabular}{|c c c c|}
\hline
X ray sources & Distance & Type & References\\
\hline
GX1+4                  & 10.0   & LMXB & \cite{grimm2002}\\
X0512-401              & 12.2   & LMXB & \cite{liu2007}\\
X0921-630              & 7.0    & LMXB & \cite{liu2007}\\
X1246-588              & 5.0    & LMXB & \cite{liu2007}\\
SAXJ1324.5-6313        & 6.2    & LMXB & \cite{liu2007}\\
X1524-617              & 4.4    & LMXB & \cite{liu2007}\\
MS1603.6+2600          & 5.0    & LMXB & \cite{liu2007}\\
X1711-339              & 7.5    & LMXB & \cite{liu2007}\\
SAXJ2224.9+5421        & 7.1    & LMXB & \cite{liu2007}\\
GRS1737-31             & 8.5    & LMXB & \cite{liu2007}\\
EXO1747-214            & 11.0   & LMXB & \cite{liu2007}\\
XTEJ1748-288           & 8.0    & LMXB & \cite{liu2007}\\
SAXJ1752.3-3128        & 9.0    & LMXB & \cite{liu2007}\\
SWIFTJ1753.5-0127      & 6.0    & LMXB & \cite{liu2007}\\
SAXJ1806.5-2215        & 8.0    & LMXB & \cite{liu2007}\\
XTEJ1817-330           & 1.0    & LMXB & \cite{liu2007}\\
SAXJ1818.7+1424        & 9.4    & LMXB & \cite{liu2007}\\
X1850-087              & 8.2    & LMXB & \cite{liu2007}\\
XTEJ1908+094           & 3.0    & LMXB & \cite{liu2007}\\
X1953+319              & 1.7    & LMXB & \cite{liu2007}\\
SAXJ0635+0533          & 2.5    & HMXB & \cite{liu2006}\\
XTEJ1739-302           & 2.3    & HMXB & \cite{liu2006}\\
AXJ1841-0536           & 10.0   & HMXB & \cite{liu2006}\\
X1901+031              & 10.0   & HMXB & \cite{liu2006}\\
XTEJ1906+09            & 10.0   & HMXB & \cite{liu2006}\\
GROJ2058+42            & 9.0    & HMXB & \cite{liu2006}\\
SAXJ2239.3+6116        & 4.4    & HMXB & \cite{liu2006}\\
HerX-1		       & 6.6	& LMXB & \cite{reynolds}\\
EXO1722-363	       & 8.0	& HMXB & \cite{bodaghee}\\
X1704+240	       & 0.42	& LMXB & \cite{liu2007}\\
RXJ1709.5-2639	       & 11.0	& LMXB & \cite{liu2007}\\
X2127+119	       & 5.8	& LMXB & \cite{galloway2008}\\
\hline
\end{tabular}
\end{table}

\end{document}